\documentstyle[pre,preprint,aps]{revtex}
\tightenlines
\begin{document}
\draft
%\baselineskip=0.5\baselineskip

\title{
Dynamical mean-field theory of spiking  
neuron ensembles: 
response to a single spike with independent noises
\footnote{E-print: cond-mat/0206135}
}
\author{
Hideo Hasegawa
\footnote{E-mail:  hasegawa@u-gakugei.ac.jp}
}
\address{
Department of Physics, Tokyo Gakugei University,
Koganei, Tokyo 184-8501, Japan
}
\date{\today}
\maketitle
\begin{abstract}
Dynamics of an ensemble of $N$-unit FitzHugh-Nagumo (FN) neurons
subject to white noises has been studied
by using a semi-analytical
dynamical mean-field (DMF) theory in which
the original $2 N$-dimensional 
{\it stochastic} differential equations 
are replaced by 8-dimensional {\it deterministic}
differential equations expressed in terms
of moments of local and global variables.
Our DMF theory, which assumes weak noises and the Gaussian 
distribution of state variables, goes
beyond weak couplings among constituent neurons.
By using the expression for
the firing probability due to an applied single spike,
we have discussed
effects of noises, synaptic couplings
and the size of the ensemble
on the spike timing precision, which is shown to be 
improved by increasing the size of the neuron ensemble,
even when there are no couplings among neurons.
When the coupling is introduced, neurons in ensembles
respond to an input spike with a partial
synchronization. 
DMF theory is extended to a large cluster which can be 
divided into multiple sub-clusters according to their functions.
A model calculation has shown that
when the noise intensity is moderate,
the spike propagation with a fairly precise timing is possible 
among noisy sub-clusters with feed-forward couplings,
as in the synfire chain. 
Results calculated by our DMF theory
are nicely compared to those obtained by direct simulations.
A comparison of DMF theory with the conventional moment
method is also discussed.

\end{abstract}

\noindent
\vspace{0.5cm}
%
%\narrowtext
\section{INTRODUCTION}

%\begin{center}
%{\bf I. INTRODUCTION}
%\end{center}

It has been controversial how neurons communicate 
information by firings or spikes \cite{Rieke96}-\cite{Pouget00}.
Much of debates on the nature of the neural code has been
mainly focused on the two issues.
The first issue is whether information is encoded
in the average firing rate of neurons ({\it rate code})
or in the precise firing times ({\it temporal code}).
%It has been widely believed that information is encoded
%in the average firing rate of individual neurons ({\it rate code}).
Adrian \cite{Adrian26} first noted the relationship
between the neural firing rate and the stimulus intensity, which
forms the basis of the rate code.
Actually firing activities of motor and sensory neurons are
reported to vary in response to applied stimuli.
In recent years, however, an alternative 
temporal code has been proposed in which detailed
spike timings are assumed to play an important role in information
transmission:
information is encoded in interspike
intervals or in relative timings between
firing times of spikes \cite{Softky93}-\cite{Stevens98}.
Indeed, experimental evidences have accumulated 
in the last several years, indicating a  use of
the temporal coding in neural systems \cite{Carr86}-\cite{Thorpe96}.
Human visual systems, for example, have shown to classify
patterns within 250 ms despite the fact that at least 
ten synaptic stages are involved from retina to 
the temporal brain \cite{Thorpe96}.
The transmission
times between
two successive stages of synaptic 
transmission are suggested to be no more than 10 ms 
on the average.
This period is too short to allow rates to be determined
accurately.

The second issue is whether
information is encoded in the activity of a single (or very few) neuron
or that of a large number of neurons
({\it population} or {\it ensemble code}).
The population rate-code model assumes that information is
coded in the relative firing rates of ensemble neurons,
and has been adopted in the most of the theoretical analysis
\cite{Abbott98}.
On the contrary, in the population temporal-code model,
it is assumed that
relative timings between spikes in ensemble neurons
may be used as an encoding mechanism for perceptional
processing \cite{Hopfield95}-\cite{Rullen01}.
A number of experimental data supporting this code have been reported
in recent years \cite{Gray89}-\cite{Hatso98}.
For example, data has demonstrated that temporally
coordinated spikes can systematically signal sensory
object feature, even in the absence of changes
in firing rate of the spikes \cite{deCharms96}.

It is well known that neurons in brains are subject
to various kinds of noises, which can alter
the response of neurons in various ways.
Although firings of a single neocortical 
neuron in {\it vitro} are precise and
reliable, those in {\it vivo} are quite unreliable \cite{Mainen95}.
This is due to noisy environment in {\it viro},
which makes the reliability of 
neurons firings worse.
The strong criticism against the temporal code is that
spikes are vulnerable to noise while the rate code
performs robustly in the presence of noise but with
limited information capacity.
It has been shown, however, that
the response of neurons is improved by 
background noises
against our conventional wisdom.
The typical example is the stochastic resonance (SR),
in which weak noises enhance 
the transmission of subthreshold signals
(for review see Refs. \cite{Gammai98}\cite{Anish99}).
It has been shown that noise of appropriate magnitude
linearlizes the response of neurons,
which leads to SR and maximizes input-output correlation,
transformation and coherence 
(for review see Ref.\cite{Segund94}).
Recently, 
it has been demonstrated that 
noises can enhance the firing-time reliability
of neurons stimulated by weak periodic and aperiodic inputs
\cite{Tanabe99}-\cite{Tanabe01b}.
We may expect that a population of neuron ensembles
plays important roles in the response of neurons
subject to noises.
Actually SR in HH neuron ensembles has first demonstrated
for a single input
by Pei, Wilkens and Moss \cite{Pei96a}.
Subsequently this pooling effect has been supported 
for aperiodic \cite{Collins95a}\cite{Chialvo97}
and periodic (analog) signals \cite{Shimokawa99}
and for spike-train inputs
\cite{Hasegawa02a}\cite{Hasegawa02b}.
Quite recently,
SR for a transient spike signal 
in large-scale HH neuron ensembles has been studied
by using the wavelet analysis {\cite{Hasegawa02a} .
It may be possible that the firing-time precision is also 
improved by increasing the size of neuron ensembles.

A small patch of cortex may contain thousands 
of similar neurons, each connecting with hundreds or thousands of 
other neurons in that same patch or in other patches.
%It is well known that a distinct cortical area contains $O\:(10^8)$
%neurons and a cortical neuron interacts with as many as $O\:(10^5)$
%other neurons in the same area or in the other areas.
The underlying dynamics of individual neurons includes a variety of 
voltage dependent ionic channels which can be described by 
Hodgkin-Huxley-type differential equations.
Computational neuroscientists have so far 
tried to gain understanding of the
properties of neuron ensembles
with the use of two approaches: direct simulations
and mean-field (MF) theories.
Simulations have been made for large-scale
networks mostly consisting of integrate-and-fire (IF) neurons.
Since the time to simulate networks by conventional methods 
grows as $N^2$ 
with $N$, the size of the network \cite{Hansel98},
it is rather difficult to simulate realistic neuron clusters,
in spite of recent computer development.  
In MF theories \cite{Kuramoto91}-\cite{Gerstner95}, 
dynamics of globally
coupled large-scale networks is described by a flow of phases
or the population activity,
which determines the fraction of the firing rate of neurons.
The stability condition for synchronous and asynchronous states
of neuron clusters has been investigated.
Quite recently, the population density method has been developed
as a tool modeling for large-scale neuronal clusters 
\cite{Omurtag00}-\cite{Haskell00}. 
In these MF approaches the macroscopic variable of interest
is the firing rate, following the {\it rate-code} hypothesis.
However, only little MF approaches have so far proposed based
on the temporal-code hypothesis \cite{Treves93}\cite{Gerstner95}.

The purpose of the present study is to construct 
a dynamical mean-field (DMF) theory based 
on the temporal code hypothesis,
generalizing the method previously proposed by
Rodriguez and Tuckwell (RT) \cite{Rod96}-\cite{Rod00}.
In the RT theory, the dynamics of the membrane potential
of a neuron subject to white noises
is studied by replacing {\it stochastic} differential
equations (DEs) by {\it deterministic} DEs
described by moments
of state variables.
RT's general theory has first applied to a single 
FitzHugh-Nagumo (FN) neuron 
\cite{Rod96}\cite{Tuckwell98}
and then a Hodgkin-Huxley (HH) neuron \cite{Rod98}\cite{Rod00}.
In the case of a single FN neuron, for example, 
two stochastic DEs 
are replaced by five deterministic DEs,
for which an improvement to the RT theory has been recently
proposed \cite{Tanabe01}.
When the RT theory is applied to a $N$-unit FN neuron network, 
$2 N$ stochastic DEs are replaced by $N_{eq}=N\;(2N+3)$ 
deterministic DEs \cite{Rod96}\cite{Rod00}.
For example, in the case of $N=100$, we get $N_{eq}=20300$, 
which is too large to perform calculations for neuron ensembles. 
In their subsequent paper \cite{Rod98},
the result of ensemble neurons is transplanted to 
the Fokker-Planck (FP) equation 
for the transition probability density,
which is a partial differential equation with $2N+1$ independent
variables. Solving such a FP equation is a hard computational task. 
%Solving $5 N$ DEs is, however, laborious when the size of the cluster, $N$,
%becomes large, although deterministic DEs is more amenable
%than stochastic DEs.
We will present in this paper, an alternative MF
theory for $N$ FN neuron ensembles, 
replacing original
$2 N$ stochastic DEs by {\it eight} deterministic DEs 
which are expressed in terms of 
means, variances and covariances 
of {\it local} and {\it global} state variables. 

There are several nonlinear models which have been
used for a study of neuron activities.
Among them we employ here the FN model 
\cite{FitzHugh61}\cite{Nagumo62}
because
it is relatively simple and amenable to analysis
although the FN model does not have as firm an empirical
basis as conductance-based model like the HH model.
The property of the FN model has been intensively 
investigated.
In recent years, SR of a single FN neuron \cite{Longtin93}-\cite{Longtin94}
and FN neuron ensembles 
\cite{Collins95a}\cite{Chialvo97}\cite{Kanamaru01}\cite{Stocks01} 
have been studied.

The paper is organized as follows:
In Sec. II, we have developed a DMF theory
for $N$ FN neuron ensembles, expanding the original
stochastic DEs in terms of deviations from means to get
variances and covariances of local and global variables.
We compare our DMF theory with conventional RT's 
moment method \cite{Rod96}, showing that the former may
be derived from the latter.
Some calculated results are reported of the response of 
ensemble neurons to a single spike with white noises.
It will be shown that the spike firing precision is 
improved by increasing the ensemble
size and the synaptic couplings, as expected.
In Sec. III.  DMF theory is extended to
a large cluster consisting of multiple sub-clusters
and model calculations are reported.
The final Sec. IV is devoted to
conclusions and discussions.

\section{Neuron Ensembles}
\subsection{DMF approximation}

We assume a neuron ensemble consisting of $N$-unit
FN neurons.
Dynamics of a single FN neuron $i$ in a given ensemble
is described by the nonlinear DEs given by 
\begin{eqnarray}
\frac{dx_{i}(t)}{dt} &=& F[x_{i}(t)]
- c y_{i}(t) + I_{i}^{(c)}(t) 
+I^{(e)}(t) + \xi_i(t), \\
\frac{dy_{i}(t)}{dt} &=& b x_{i}(t) - d y_{i}(t)+e,
\;\;\;\mbox{($i=1-N$)}
\end{eqnarray}
where $F[x(t)]=k\: x(t)\: [x(t)-a]\: [1-x(t)]$, 
$k=0.5$, $a=0.1$, $b=0.015$, $c=1.0$, $d=0.003$ and $e=0$ 
\cite{Rod96}\cite{Tuckwell98},
$x_i$ and $y_i$ denote the fast (voltage) variable
and slow (recovery) variable, respectively,
and $\xi_i(t)$ is the independent
Gaussian white noise with $<\xi_i(t)>=0$ and
$<\xi_i(t)\:\xi_j(t')>=\beta_i^2 \; \delta_{ij}\:\delta(t-t')$,
the bracket $< \cdot >$ denoting the average
over stochastic random variables \cite{note1}.
In Eq. (1), 
$I_i^{(c)}(t)$ denotes the coupling term given by
\begin{equation}
I_i^{(c)}(t)=\frac{w}{N}\;\sum_{j (\neq i)} G(x_{j}(t)),
\end{equation}
where 
$w$ stands for the coupling strength and 
$G(x)=1/[1+{\rm exp}[-(x-\theta)/\alpha]]$ is the sigmoid function
with the threshold $\theta$ and the width $\alpha$.
The self-coupling terms are excluded in Eq. (3), where we have adopted
the normalization factor to be $N^{-1}$
instead of $(N-1)^{-1}$ for a later study of the $N=1$ limit.
$I^{(e)}(t)$ expresses an external, single spike input 
applied to all neurons, given by
\begin{equation}
I^{(e)}(t)
%=I_{in}(t) 
=A \;\Theta(t-t_{in})\; \Theta(t_{in}+T_w-t),
\end{equation}
where $\Theta(x)=1$ for $x>0$ and 0 otherwise,  
$A$ stands for the magnitude, $t_{in}$ the input time and 
$T_w$ the width.

After RT \cite{Rod96}\cite{Tuckwell98},
we will express these nonlinear DEs
by moments of variables.
First we define the global variables for the ensemble by
\begin{eqnarray}
X(t)&=&(1/N)\;\sum_{i} \;x_{i}(t), \\
Y(t)&=&(1/N)\;\sum_{i} \;y_{i}(t),
\end{eqnarray}
and their averages by
\begin{eqnarray}
\mu_1(t)&=&<X(t)>,  \\ 
\mu_2(t)&=&<Y(t)>.
\end{eqnarray}
Next we express the differential equations given by 
Eqs. (1) and (2) in terms
of the deviations from their averages defined by
\begin{eqnarray}
\delta x_i(t)&=& x_i(t)-\mu_1(t), \\
\delta y_i(t)&=& y_i(t)-\mu_2(t),
\end{eqnarray}
to get (the argument $t$ is hereafter neglected) 
\begin{eqnarray}
\frac{d x_i}{d t} &=& F(\mu_1) + F'(\mu_1) \delta x_i 
+\frac{1}{2} F^{(2)}(\mu_1) \delta x_i^2 
+\frac{1}{6} F^{(3)}(\mu_1) \delta x_i^3 \nonumber \\
&& -c \mu_2 -c \delta y_i + I_i^{(c)} 
+ I_i^{(e)} + \xi_i, \\
\frac{d y_i}{d t} &=& b \mu_1 - d \mu_2 
+b \delta x_i - d \delta y_i+e,
\end{eqnarray}
with
\begin{eqnarray}
I_i^{(c)}&=& w \{ (1-\frac{1}{N}) G(\mu_1) \nonumber \\
&&+ \frac{1}{N} \sum_{j (\neq i)} [G'(\mu_1) \delta x_j 
+ \frac{1}{2} G^{(2)}(\mu_1) \delta x_j^2
+ \frac{1}{6} G^{(3)}(\mu_1) \delta x_j^3]).
\end{eqnarray}
We define the variances and covariances
between local variables, given by 
\begin{eqnarray}
\gamma_{1,1}&=& \frac{1}{N}\; \sum_{i} <\delta x_i^2>, \\
\gamma_{2,2}&=& \frac{1}{N}\; \sum_{i} <\delta y_i^2>, \\
\gamma_{1,2}&=& \frac{1}{N}\; \sum_{i} 
<\delta x_i \;\delta y_i>,
\end{eqnarray}
and those between global variables, given by
\begin{eqnarray}
\rho_{1,1}&=& <\delta X^2>, \\
\rho_{2,2}&=& <\delta Y^2>, \\
\rho_{1,2}&=& <\delta X \;\delta Y>,  
\end{eqnarray}
where $\delta X = X(t)-\mu_1(t)$ and $\delta Y = Y(t)-\mu_2(t)$.
It is noted that $\gamma_{\kappa,\lambda}$ expresses the
spatial average of fluctuations
in local variables of $x_i$ and $y_i$ 
while $\rho_{\kappa,\lambda}$
denotes fluctuations in global variables of $X$ and $Y$.
We assume that (1) the noise intensity $\beta$ is weak 
%(2) coupling $w$ is small and
and (2) the distribution
of state variables take Gaussian form.
The first assumption allows us to expand
the quantities in a power series of fluctuation moments
around means.
As for the second assumption, numerical simulations
have shown that for weak noises, the distribution of $x(t)$ of the 
membrane potential of a single FN neuron nearly obeys
the Gaussian distribution,
although for strong noises, the distribution of $x(t)$
deviates from the Gaussian, taking a bimodal form
(see Fig. 8 of Ref.\cite{Tuckwell98} and Fig. 3 of Ref.\cite{Tanabe01}).
Similar behavior of the membrane-potential distribution 
has been reported in a HH neuron model \cite{Shimokawa99}\cite{Tanabe01a}.
When adopting the Gaussian assumption, we may express
the average of fluctuations in terms of the
first and second moments only.
It is noted that we impose no conditions on the
coupling strength.
After some manipulations, we get DEs for means,
variances and covariances, given by
(for details see Appendix A):

\begin{eqnarray}
\frac{d \mu_1}{d t}&=&f_0 + f_2 \gamma_{1,1} -c \mu_2 
+ w\;(1-\frac{1}{N})\;U_0+I^{(e)}(t), \\
\frac{d \mu_2}{d t}&=& b \mu_1 - d \mu_2 +e
,  \\
\frac{d \gamma_{1,1}}{d t}&=& 2 [(f_1+ 3 f_3 \gamma_{1,1} ) 
\gamma_{1,1}- c \gamma_{1,2}] 
+ 2 w (\rho_{1,1}-\frac{\gamma_{1,1}}{N})\;U_1 +\beta^2, \\
\frac{d \gamma_{2,2}}{d t}&=& 2 (b \gamma_{1,2}- d \gamma_{2,2}),  \\
\frac{d \gamma_{1,2}}{d t}&=& b \gamma_{1,1}+ (f_1+3 f_3 \gamma_{1,1}-d) 
\gamma_{1,2} 
- c \gamma_{2,2}
+ w\;(\rho_{1,2}-\frac{\gamma_{1,2}}{N})\;U_1,  \\
\frac{d \rho_{1,1}}{d t}&=& 2 [(f_1+ 3 f_3 \gamma_{1,1} )\rho_{1,1} 
- c \rho_{1,2}]
+ 2 w \;(1-\frac{1}{N})\; \rho_{1,1}\;U_1
+ \frac{\beta^2}{N}, \\
\frac{d \rho_{2,2}}{d t}&=& 2 (b \rho_{1,2}- d \rho_{2,2}),  \\
\frac{d \rho_{1,2}}{d t}&=& b \rho_{1,1}+ (f_1 +3 f_3 \gamma_{1,1}-d) \rho_{1,2} 
- c \rho_{2,2}
+ w \;(1-\frac{1}{N})\;\rho_{1,2}\;U_1,  
\end{eqnarray}
with
\begin{eqnarray}
U_0&=&g_o+g_2 \gamma_{1,1}, \\
U_1&=&g_1+3g_3 \gamma_{1,1}, \\
f_{\ell} &=& (1/\ell !) F^{(\ell)}(\mu_1), \\
g_{\ell} &=& (1/\ell !) G^{(\ell)}(\mu_1).
\end{eqnarray}
where $\beta^2 = (1/N) \sum_i \beta_i^2$.

The original $2 N$-dimensional {\it stochastic} differential
equations given by Eqs. (1) and (2) are transformed to
eight-dimensional {\it deterministic} differential equations,
which show much variety depending on model parameters such
as the strength of white noise ($\beta$), 
couplings ($w$) and
the size of cluster ($N$).

\subsection{Derivation of DMF theory from RT's moment method}

Before discussing the property of our DMF theory,
we will show that it can be derived from 
RT's moment method \cite{Rod96}.
In the case of a single FN neuron $(N=1)$, DMF theory agrees
with the RT theory as shown in appendix B,
where some limiting cases
of Eqs. (20)-(27) are examined.
In the case of FN neuron ensemble ($N \geq 2$),
the RT theory defines means of variables 
of $x_i$ and $y_i$
for the neuron $i$ given by
\begin{eqnarray}
m_1^i &=& <x_i>, \\
m_2^i &=& <y_i>,
\end{eqnarray}
and calculate variances and covariances between
{\it local} variables as given by
\begin{eqnarray}
C_{1,1}^{i,j} &=& <\Delta x_i \; \Delta x_j >, \\
C_{2,2}^{i,j} &=& <\Delta y_i \: \Delta y_j >, \\
C_{1,2}^{i,j} &=& <\Delta x_i \; \Delta y_j>,
\end{eqnarray}
where $\Delta x_i=x_i-m_1^i$ and $\Delta y_i= y_i-m_2^i$.
Variances are given by setting $i=j$ in Eqs. (34)-(35).
Adopting the same assumptions as our DMF theory:
(1) weak noises and (2) the Gaussian distribution 
for state variables,
we get DEs for these moments given by
(for details see Appendix A)
%; see also Eqs. (21a)-(23b) of Ref.\cite{Rod96})
\begin{eqnarray}
\frac{d m_1^i}{d t}&=&f_0^i + f_2^i C_{1,1}^{i,i} -c m_2^i 
+ \frac{w}{N} \sum_{j(\neq i)} 
[g_o^j+g_2^j C_{1,1}^{j,j}]+I^{(e)}, \\
\frac{d m_2^i}{d t}&=& b m_1^i - d m_2^i +e,  \\
\frac{d C_{1,1}^{i,j}}{d t}&=&  (f_1^i+ 3 f_3^i C_{1,1}^{i,i}
+f_1^j+ 3 f_3^j C_{1,1}^{j,j}) C_{1,1}^{i,j} 
- c (C_{1,2}^{i,j}+C_{2,1}^{i,j})
+\beta^2 \delta_{ij}  \nonumber \\
&&+ \frac{w}{N} [\sum_{k(\neq i)} 
(g_1^k+3 g_3^k C_{1,1}^{k,k}) C_{1,1}^{j,k}
+ \sum_{k(\neq j)} 
(g_1^k+3 g_3^k C_{1,1}^{k,k}) C_{1,1}^{i,k}], \\
\frac{d C_{2,2}^{i,j}}{d t}&=& b (C_{1,2}^{i,j}+C_{2,1}^{i,j})
- 2 d C_{2,2}^{i,j},  \\
\frac{d C_{1,2}^{i,j}}{d t}&=& b C_{1,1}^{i,j}
+ (f_1^i+3 f_3^i C_{1,1}^{i,i}-d) C_{1,2}^{i,j}-c C_{2,2}^{i,j} 
\nonumber \\
&&+ \frac{w}{N} \sum_{k(\neq i)} 
(g_1^k+3 g_3^k C_{1,1}^{k,k}) C_{2,1}^{j,k},
\hspace{3cm} \mbox{($i,j = 1$ to $N$)}
\end{eqnarray}
where
\begin{eqnarray}
f_{\ell}^i &=& (1/\ell !) F^{(\ell)}(m_1^i), \\
g_{\ell}^i &=& (1/\ell !) G^{(\ell)}(m_1^i).
\end{eqnarray}
%($g_{\ell}^i$ is set zero for $\ell \geq 3$ for
%a simplicity of discussion).

Now we derive DMF theory from RT's moment method.
We adopt the approximation given by
\begin{equation}
m_{1}^i=\mu_{1},
\end{equation}
which yields $f_{\ell}^i=f_{\ell}$ and $g_{\ell}^i=g_{\ell}$
[Eqs. (28), (29), (42) and (43)],
and the approximation given by
\begin{eqnarray}
f_3 C_{1,1}^{i,i} &\simeq& f_3 \gamma_{1,1}, \nonumber \\
g_3 C_{1,1}^{k,k} &\simeq& g_3 \gamma_{1,1}.
\end{eqnarray}
in Eqs. (39) and (40).
We realize that
quantities of $\mu_{\kappa}$, $\gamma_{\kappa,\lambda}$ 
and $\rho_{\kappa,\lambda}$ [Eqs. (5)-(8) and (14)-(19)]
adopted in DMF theory are expressed in terms of $m_{\kappa}^i$ 
and $C_{\kappa,\lambda}^{i,j}$ [Eqs. (32)-(36)]
in the RT theory as follows:
\begin{eqnarray}
\mu_{\kappa}&=& \frac{1}{N} \sum_{i} m_{\kappa}^i, \\
\gamma_{\kappa,\lambda}&=& \frac{1}{N} \sum_{i} C_{\kappa,\lambda}^{i,i}, \\
\rho_{\kappa,\lambda}&=& \frac{1}{N^2} \sum_{i} \sum_{j} 
C_{\kappa,\lambda}^{i,j}.
\hspace{2cm} \mbox{($\kappa, \lambda=1,2$)}
\end{eqnarray}
Then, 
we may obtain,
from Eqs. (37)-(41), (44)-(48), 
alternative DEs for 
$\mu_{\kappa}$, $\gamma_{\kappa,\lambda}$
and $\rho_{\kappa,\lambda}$
which are again given by Eqs. (20)-(27).
This implies that 
DMF theory may be derived
from RT's moment method if we adopt the assumptions given by Eq. (44)
and (45) \cite{note3}.

Taking into the symmetry relations: 
$C_{\kappa,\lambda}^{i,j}=C_{\kappa,\lambda}^{j,i}$, we get 
the number of DEs to be $N_{eq}\;= \;2 N+N(2N+1)\;=\;N(2N+3)$
in the RT theory [Eqs. (37)-(41)] 
while $N_{eq}=8$ in our DMF theory [Eq. (20)-(27)]
($N_{eq}= 2N N_{tr}$ in direct simulations 
where $N_{tr}$ denotes the number of trials). 
In the case of $N=100$, for example,
we get $N_{eq}$=20300 in the RT method, which is 
much larger than $N_{eq}=8$ in DMF theory.
Our DMF theory successfully reduces the number of DEs,
by taking account
$\mu_{\kappa}$, $\gamma_{\kappa,\lambda}$ 
and $\rho_{\kappa,\lambda}$ for
{\it global} variables as well as {\it local} variables
instead of $m_{\kappa}^i$ and
$C_{\kappa,\lambda}^{i,j}$ for local variables.
Although our DMF theory neglects spatial fluctuations
in state variables, it has 
advantages of a tractable small number of DEs and clear
semi-analytical nature, from which some qualitative
results may be deduced without numerical calculations,
as will be shown shortly
[{\it e.g.} Eq. (52)].
When couplings have the spatial dependence as $w \rightarrow w_{ij}$
in Eq. (3), we have to rely on Eqs. (37)-(41) in the moment method.

\subsection{Property of DMF theory}

In this subsection, we will discuss the property of our DMF theory.
It is possible
to regard DMF theory as the single-site mean-field theory.
Let us assume a configuration in which a {\it single}
neuron $i$ is embedded in an effective {\it medium} whose effect
is realized by a given neuron $i$
as its {\it effective } external input
through the coupling $w$.
We replace quantities of $m_{\kappa}^k$, 
$C_{\kappa,\lambda}^{k,k}$ and
$(1/N) \sum_{k(\neq \ell)} 
C_{\kappa,\lambda}^{\ell,k}$ in coupling
terms of Eqs. (37)-(41) by effective quantities of $\mu_{\kappa}$, 
$\gamma_{\kappa,\lambda}$ and
$\rho_{\kappa,\lambda}-(1/N) \gamma_{\kappa,\lambda}$,
respectively.
Then, in order to determine these quantities
just introduced, we impose the 
{\it single-site} mean-field conditions given by [see Eqs. (46)-(48)]
\begin{eqnarray}
\mu_{\kappa}&=& m_{\kappa}^i, \\
%(=\frac{1}{N} \sum_i m_{\kappa}^i), \\
\gamma_{\kappa,\lambda}&=& C_{\kappa,\lambda}^{i,i}, \\
%(= \frac{1}{N} \sum_i C_{\kappa,\lambda}^{i,i}), \\
\rho_{\kappa,\lambda}-\frac{1}{N} \gamma_{\kappa,\lambda} 
&=& \frac{1}{N} \sum_{j (\neq i)} C_{\kappa,\lambda}^{i,j}.
%(= \frac{1}{N^2} \sum_i \sum_{j (\neq i)} C_{\kappa,\lambda}^{i,j}).
\end{eqnarray}
Note that Eqs. (49)-(51) are assumed to hold independently of $i$ and
that $m_{\kappa}^i$
and $C_{\kappa,\lambda}^{i,j}$ in their righthand sides 
are functions of
$\mu_{\kappa}$, $\gamma_{\kappa,\lambda}$ 
and $\rho_{\kappa,\lambda}$.
Conditions given by Eqs. (49)-(51) 
yield the self-consistent DEs for
$\mu_{\kappa}$, $\gamma_{\kappa,\lambda}$
and $\rho_{\kappa,\lambda}$
which are again given by Eqs. (20)-(27).
The single-site approximation given by Eq. (49)-(51), which 
assumes that the quantities averaged at a given site are the same as
those of the effective medium, is common in mean--field theories
such as the Weiss theory for magnetism \cite{Weiss07}
%BCS theory for superconductivity 
and the coherent-potential 
approximation for random alloys \cite{Soven67}.

We should note that
the noise contribution is $\beta^2$ in Eq. (22)
while that is $\beta^2/N$ in Eq. (25).
It is easy to see that in the case of no couplings,
we get
\begin{equation}
\rho_{\kappa, \lambda}=\frac{\gamma_{\kappa, \lambda}}{N},
\hspace{2cm}\mbox{(for $w=0$)}
\end{equation}
which agrees with the {\it central-limit theorem}.
%, and
%which is the origin of the pooling effect \cite{deCharms00}.
On the other hand, in the case of $\beta=0$ and $w \neq 0$,
we get $\rho_{\kappa, \lambda}=\gamma_{\kappa, \lambda}$.
Thus the ratio: $\rho_{\kappa, \lambda}/\gamma_{\kappa, \lambda}$ 
changes as model parameters are changed.
We will show that these changes in 
$\rho_{\kappa, \lambda}$and $\gamma_{\kappa, \lambda}$
reflect on the firing time distribution and the degree of
synchronous firings in neurons ensembles.

\vspace{0.5cm}
\noindent
{\bf Firing-Time Distribution}

The ($n$th) firing time of a given neuron $i$ in the
ensemble is defined as the time when $x_i(t)$ 
crosses the threshold $\theta$ from below:
\begin{equation}
t_{oin}= 
\{ t \mid x_i(t) = \theta; \dot{x_i} >0;
t \geq t_{oin-1}+\tau_r \},
\end{equation} 
where $\tau_r$ denotes the refractory period introduced so
as to avoid multiple firings in a short period
arising from fluctuations in voltage variables around 
the threshold.
%By integrating the multi-variate normal density 
%over $x_j\;(j \neq i)$ and $y_j$ ($j=1$ to $N$), 
We get the distribution for the
membrane potential variable $x_i$ given by (for details see Appendix C)
\begin{eqnarray}
P(x_i) 
%&=& \int ... \int \; \Pi_{j(\neq i)} \:dx_j \:\Pi_j \:dy_j\;
%p(x_1, ...,x_N, y_1, ...,y_N), 
&\simeq&(\frac{1}{\sigma_{\ell}})\;\phi(\frac{x_i-\mu_1}{\sigma_{\ell}}),
\end{eqnarray} 
where $\phi(x)$ is the normal distribution function given by
\begin{equation}
\phi(x)=\frac{1}{\sqrt{2 \pi}} 
{\rm exp}(-\frac{x^2}{2}),
\end{equation} 
with
\begin{equation}
\sigma_{\ell}=\sqrt{\gamma_{1,1}}.
\end{equation}
This implies that the distribution of 
the voltage variable $x_i(t)$
is described by the Gaussian distribution with the mean
of $\mu_1(t)$ and the variance of $\gamma_{1,1}(t)$.
The probability given by Eq. (54) depends on the time because
of the time dependence of $x_i(t)$ and $\sigma_{\ell}(t)$.
The probability $W_{oi}(t)$ when $x_i(t)$ at $t$
is above the threshold $\theta$ is given by \cite{Rod98}
\begin{equation}
W_{oi}(t)=1 - \psi(\frac{\theta-\mu_1}{\sigma_{\ell}}),
\end{equation} 
where $\psi(y)$ is the error function given by integrating $\phi(x)$
from $-\infty$ to $y$.
Then the probability averaged over the ensemble is given by
\begin{eqnarray}
W_{\ell}(t)&=&\frac{1}{N}\; \sum_{i} W_{oi}(t), \nonumber \\
&=&\;1 - \psi(\frac{\theta-\mu_1}{\sigma_{\ell}}).
\end{eqnarray} 
The fraction of a given neuron $i$ emitting 
output spikes at $t$ is given by
\begin{equation}
Z_{\ell}(t)=\frac{d\;W_{o}(t)}{dt}\; \Theta(\dot{\mu_1})
= \phi(\frac{\theta-\mu_1}{\sigma_{\ell}}) \;
\frac{d}{dt}(\frac{\mu_1}{\sigma_{\ell}}) \; \Theta(\dot{\mu_1}).
\end{equation} 
where $\dot{\mu_1}=d\mu_1/dt$.
When we expand $\mu_1(t)$ in Eq. (59) around $t^{*}_{o}$ where
$\mu_1(t^{*}_{o})=\theta$,
it becomes
\begin{equation}
Z_{\ell}(t) \sim 
\phi(\frac{t-t^{*}_{o}}{\delta t_{o\ell}})\;
\frac{d}{dt}(\frac{\mu_1}{\sigma_{\ell}}) \; \Theta(\dot{\mu_1}),
\end{equation} 
with
\begin{equation}
\delta t_{o\ell}=\frac{\sigma_{\ell}}{\dot{\mu_1}},
\end{equation} 
where $\mu_1$, $\dot{\mu_1}$ and $\sigma_{\ell}$ are evaluated at $t=t^{*}_o$.
$Z_{\ell}(t)$ provides the
distribution of firing times,
showing that most of firing times of neurons locate in the range given
as
\begin{equation}
t_{o} \in [t^{*}_o-\delta t_{o\ell},\; t^{*}_o+\delta t_{o\ell}].
\end{equation} 
In the limit of vanishing $\beta$, Eq. (60) reduces to
\begin{equation}
Z_{\ell}(t)= \delta(t-t^{*}_{o}).
\end{equation} 

Similarly, 
%when we 
%integrate the bivariate normal density $p(X,Y)$
%over the variable $Y$, 
we get the distribution  for the
global variable $X$ given by (for details see Appendix D)
\begin{eqnarray}
P(X) 
%&=& \int dY  \; p(X,Y), \\
&\simeq& (\frac{1}{\sigma_g})\; \phi(\frac{X-\mu_1}{\sigma_{g}}),
\end{eqnarray} 
with
\begin{equation}
\sigma_g=\sqrt{\rho_{1,1}}.
\end{equation} 
This implies that the distribution of 
global voltage variable $X(t)$
is described by the Gaussian distribution with the mean
of $\mu_1(t)$ and the variance of $\rho_{1,1}(t)$.
If we define the $m$th firing time relevant to 
the global variable $X(t)$ as
\begin{equation}
t_{gm}=\{ t \mid X(t) = \theta; \dot{X}(t) > 0;
t \geq t_{gm-1}+\tau_r \},
\end{equation} 
the fraction of firing around $t=t^{*}_o$
is given by
\begin{equation}
Z_{g}(t)= 
\phi(\frac{t-t^{*}_o}{\delta t_{og}})\;
\frac{d}{dt}(\frac{\mu_1}{\sigma_g}) \; \Theta(\dot{\mu_1}),
\end{equation} 
with
\begin{equation}
\delta t_{og}=\frac{\sigma_g}{\dot{\mu_1}}.
\end{equation} 
Then most of $t_{og}$ locate in the range given by
\begin{equation}
t_{g} \in [t^{*}_o-\delta t_{og},\; t^{*}_o+\delta t_{og}].
\end{equation} 
Since $\rho_{1,1}$ is generally smaller than $\gamma_{1,1}$, we get
$\sigma_g \leq \sigma_{\ell}$ and $\delta t_{0g} \leq \delta t_{o\ell}$.
In particular, in the case of no couplings, 
Eqs. (52), (56), (61), (65) and (68) 
lead to
\begin{equation}
\delta t_{og}=\frac{\delta t_{o\ell}}{\sqrt{N}}.  
\hspace{2cm} \mbox{(for $w=0$)}
\end{equation}

\vspace{0.5cm}
\noindent
{\bf Synchronous Response}

Now we consider the quantity given by
\begin{equation}
R(t)=\frac{1}{N^2} \sum_{i j}
<[x_i(t)-x_j(t)]^2>=2(\gamma_{1,1}-\rho_{1,1}).
\end{equation}
When all neurons are in the completely synchronous state,
we get $x_i(t)=X(t)$ for all $i$, and then 
$R(t)=0$.
On the contrary, in the asynchronous (random) state, we get
$R(t)=2(1-1/N)\gamma_{1,1} \equiv R_0(t)$.
Then the quantity defined by 
\begin{equation}
S(t)=1 - R(t)/R_0(t)
=\frac{(\rho_{1,1}/\gamma_{1,1}-1/N)}{(1-1/N)},
\end{equation}
is 1 for the completely synchronous state
and 0 for the asynchronous state.
We hereafter call $S(t)$ the {\it synchronization ratio},
which provides the degree of synchronous firings
in the ensemble.
We get $S=0$ for $\beta \neq 0$ with no couplings ($w=0$), 
and $S=1$ for $w \neq 0$ with no noises ($\beta=0$).

\subsection{Calculated results}

We expect that our DMF equations given by Eqs. (20)-(27) may
show bifurcation,
synchronous and asynchronous states as well as 
chaotic states. 
In this study, we pay our attention to
the response of the FN neuron ensembles
to a single spike input, 
$I^{(e)}(t)$ given by Eq. (4),
which is applied to all neurons in the ensemble.
We have adopted the parameters
of $\theta=0.5$, $\alpha=0.1$,
$\tau_r=10$, $A=0.10$, $t_{in}=100$ and $T_w=10$. 
Parameter values of $w$, $\beta$
and $N$ will be explained shortly.
We get the critical magnitude of $A_c=0.0442$ below which
firings of neuron  defined by Eq. (53) cannot take place 
without noises ($\beta=0$). 
We have adopted the value of $A=0.10\; (> \:A_c)$
for a study of the response to a supra-threshold input,
related discussion being given in Sec. IV.

DMF calculations have been made by solving
Eqs. (20)-(27) by 
the forth-order Runge-Kutta method with
a time step of 0.01.
Direct simulations have been performed by solving
$2 N$ differential equations as given by  Eqs. (1) and (2) 
by using also
the forth-order Runge-Kutta method with
a time step of 0.01.
Simulation results are the average of 100 trials otherwise noticed.
Initial values of variables are set to be
$\mu_1=\mu_2=\gamma_{1,1}=\gamma_{1,1}=\gamma_{1,2}
=\rho_{1,1}=\rho_{2,2}=\rho_{1,2}=0$
in DMF calculations, and $x_i=y_i=0$ for $i=1$ to $N$ 
in direct simulations.
All calculated quantities are dimensionless.

The time courses of means of $\mu_1$ and $\mu_2$ calculated with
$\beta=0.01$, $w=0.0$ and $N=100$ are shown in Figs. 1(a) and 1(b),
respectively,
where solid curves denote the results of DMF
theory and dashed curves those of direct simulations.
We note that $\mu_1$ and $\mu_2$ obtained by two methods 
are in very good agreement and
they are indistinguishable.
At the bottom of Fig. 1(a) an input spike is plotted
[see also Fig.2(a)].
States of neurons in an ensemble when an input spike
is injected at $t=100$, are randomized because
noises have been already added since $t=0$.
Figures 1(c)-1(h) show the time courses
of various variances and covariances.
Agreements between the two methods are good
for $\gamma_{1,1}$, $\rho_{1,1}$, $\gamma_{1,2}$ and $\rho_{1,2}$.
There is a fairly good agreement for $\gamma_{2,2}$ and $\rho_{2,2}$. 
Comparing Figs. 1(c), 1(e) and 1(g) to 1(d), 1(f) and 1(h),
respectively, we note that the relation given by Eq. (52):
$\rho_{\kappa,\lambda}=\gamma_{\kappa,\lambda}/100$ valid for $w=0$, 
is supported by simulations:
note that results in Figs. 1(d), 1(f) and 1(h) are multiplied
by a factor of hundred.

Figures 2(a) shows a single spike input,
which is applied
at $t=100$ with a duration of $T_w=10$.
The solid curve in Fig. 2(b) express $Z_{\ell}$, 
the firing probability of the local variable $x_i(t)$,
which is a positive derivative of $W_{\ell}$
%the probability of the variable to locate
%above the threshold 
shown by the dashed curve
[Eqs. (58) and (59)].
They are calculated 
for $\beta=0.01$, $w=0.0$ and $N=100$ in DMF theory.
For a comparison, the simulation result for $Z_{\ell}$ 
is plotted in Fig. 2(c).
Firings of neurons occur at $t_o \sim$ 104 to 105
with a delay of about $4 \sim 5$. 
Fluctuations of firing times of local variables
$\delta t_{o\ell}$ are 0.37 calculated by Eq. (61) in DMF theory,
and 0.41 in simulations which is the root-mean-square (RMS)
value of firing times defined by Eq. (53).
In contrast, dashed and solid curves in Fig. 2(d) 
show $W_{g}$ and $Z_{g}$, respectively, 
for the global variable $X(t)$ in DMF theory, while Fig. 2(e)
shows $Z_{g}$ obtained in simulations.
Fluctuations in spike timings of the global variable
are $\delta t_{og}$ = 0.037 calculated by Eq. (68) in DMF theory,
and 0.041 in simulations 
which is the RMS value of firing times defined by Eq. (66).
These figures of $\delta t_{og}$ for the global variable
are ten times smaller 
than respective values of
$\delta t_{o\ell}$ for the local variable.

\vspace{0.5cm}

\noindent
{\bf Noise-strength ($\beta$) dependence}

We expect that as the noise strength is more increased, 
the distribution of membrane potentials is more widen
and fluctuations of firing times are more increased.
Filled squares in Fig. 3(a) show the $\beta$ dependence of
$\delta t_{o\ell}$ obtained by DMF theory [Eq. (61)]
with $w=0.0$ and $N=100$, while
open squares express the RMS value of firing times
obtained by simulations.
The agreement between the two methods is fairly good.
In contrast, filled circles in Fig. 3(a) show 
the $\beta$ dependence of
$\delta t_{og}$ relevant to the global variable 
obtained by DMF theory [Eq. (68)]
and open circles
stand for RMS values of firing times in simulations.
We note that $\delta t_{og}$ is much smaller
than $\delta t_{o\ell}$ although 
both $\delta t_{og}$ and $\delta t_{o\ell}$
are proportional to $\beta$ for weak noises
under consideration. 
%fluctuations of the global variable are
%much smaller than those of the local variable
%independent of $\beta$.
Figure 3(b) will be explained shortly in connection to the
result of the $w$ dependence.

\vspace{0.5cm}

\noindent
{\bf Cluster-size ($N$) dependence}

Filled squares in Fig. 4(a) show the $N$ dependence of 
the local fluctuation of
$\delta t_{o\ell}$ for $\beta=0.01$ and $w=0.0$,
obtained by DMF theory, while
open squares express that obtained by simulations.
Simulations have not
been performed for $N > 100$ because of
a limitation in our computer facility.
We note that $\delta t_{o\ell}$
is independent of $N$ because of no couplings ($w=0$).
In contrast,
filled circles in Fig. 4(a) show the $N$ dependence of
the global fluctuation of
$\delta t_{og}$ obtained by DMF theory
while open circles that by simulations.
The relation: $\delta t_{og} \propto (1/\sqrt{N})$,
holds as given by Eq. (70).
Figure 4(b) for finite $w$ will be discussed shortly.

\vspace{0.5cm}

\noindent
{\bf Coupling-strength ($w$) dependence}

So far we have neglected coupling $w$ among neurons, which is
now introduced.
Filled squares in Fig. 3(b) show the $\beta$ dependence 
of local fluctuations of
$\delta t_{o\ell}$ calculated by DMF theory 
for $w=0.2$ and $N=100$, while open squares that obtained 
by simulations. 
Filled and open circles express
global fluctuations of $\delta t_{og}$
in the DMF theory and simulations, respectively.
Comparing these results with those for $w=0.0$
shown in Fig. 3(a), we note that $\delta t_{o\ell}$ is
much reduced as $w$ is increased although there is little change 
in $\delta t_{og}$.

This is more clearly seen in Fig. 5(a),
which shows the $w$ dependence of firing-time fluctuations.
Filled squares in Fig. 5(a) show fluctuations of 
$\delta t_{o\ell}$ for the local variable
obtained for $\beta=0.01$ and $N=100$
by the DMF theory while open squares express those
calculated by simulations.
Filled and open circles in Fig. 5(a)
show fluctuations of $\delta t_{og}$ for the global variable
obtained by the DMF theory 
and simulations, respectively,
When $w$ is increased, $\delta t_{o\ell}$ is 
considerably decreased
whereas $\delta t_{og}$ is almost constant.
Figure 5(b) shows a similar plot of the $w$ dependence
of firing times when the size of an ensemble is reduced
to $N=10$. We note that $\delta t_{og}$ for $N=10$
is 3.16 times larger than $\delta t_{og}$ for
$N=100$ because $\delta t_{og}$ is proportional to
$1/\sqrt{N}$. 

Results obtained by DMF theory are analyzed in the 
Appendix E, where we get the expression
for the $w$- and $N$-dependent $\delta t_{o\ell}$
given by [see Eq. (E1)]
\begin{eqnarray}
\frac{\delta t_{o\ell}(w,N)}{\delta t_{o\ell}(0,1)}
&\sim& 1 -(\frac{1}{2}) (1-\frac{1}{N})^n\;(a_1 w  + a_2 w^2 +..), 
\end{eqnarray}
where $n=1$, $\delta t_{o\ell}(0,1)=2.71$, $a_1=7.0$
and $a_2=-11.0$.
Bold, dashed curves for $w \leq 0.2$ 
in Figs. 5(a) and 5(b) show
the $w$ dependence of $\delta t_{o\ell}$ for 
$N=100$ and 10, respectively, expressed by
Eq. (73), which are in good agreement with
results of DMF theory shown by filled squares.

Log-log plots of Fig. 4(b) show
the $N$ dependence of $\delta t_{o\ell}$ (squares)
and $\delta t_{og}$ (circles) for $w=0.2$ and $N=100$,
filled and open symbols denoting results
of DMF and simulations, respectively.
Although $\delta t_{og} \propto (1/\sqrt{N})$ as in the
case of $w=0$ [Fig. 4(a)], $\delta t_{o\ell}$ shows
the peculiar $N$ dependence, which arises from
the $(1-1/N)$ term in Eq. (73). 
The $N$ dependence given by Eq. (73) with $n=1$ and 2
is shown by thin solid curves at the uppermost in Fig. 4(b).
The result with $n=1$
is in better agreement
with the result of DMF theory shown by small filled squares
than that with $n=2$ (see Appendix E).

Couplings among neurons work
to increase the synchronous dynamics and to
suppress local fluctuations.
Figures 6(a) and 6(b) show the time sequence of the synchronization
ratio $S(t)$ defined by Eq. (72) for $w=0.1$ and $w=0.2$,
respectively, with $\beta=0.01$ and $N=100$.
Solid and dashed curves in Figs. 6(a) and 6(b) show 
results in DMF theory and simulations, respectively.
Both results are in fairly good agreement.
We note that $S(t)$ has two peaks at
times when $\rho_{1,1}(t)$ also has double peaks [Fig.1(d)].
The maximum value of $S(t)$ for $w=0.2$ is $S_{\rm max}$=0.132,
which is larger than $S_{\rm max}$=0.041 for $w=0.1$.
This trend is more clearly seen in Fig. 7 where
the maximum magnitude of $S(t)$, 
$S_{\rm max}$, is plotted as a function of $w$
for $N=10$, 20, 50 and 100.
It is shown that $S_{max}$ is increased
as the coupling strength is increased, as expected.
Figure 7 also shows that the effect of coupling is more
significant in ensembles with smaller $N$. 

An analysis of the result obtained in DMF theory
yields the expression for $w$- and $N$-dependent
$S_{max}$ given by [see Eq. (E8) in Appendix E]
\begin{equation}
S_{max}(w,N)=c_1 w + c_2 w^2 +..,
%(\frac{1}{N})(1-\frac{1}{N})
%\{ a_1 w + [b_2 + (1-\frac{1}{N})^2 b_1^2]\:w^2
%\}
\end{equation}
with 
\begin{eqnarray}
c_1&=&(\frac{1}{N})(1-\frac{1}{N})\:b_1,  \\
c_2&=&(\frac{1}{N})(1-\frac{1}{N})\:
[b_2 + (1-\frac{1}{N})^2\: b_1^2],
\end{eqnarray}
where $b_1=22$ and $b_2=-290$.
Bold, dashed curves for $w \leq 0.2$ in Fig. 7 show
$S_{max}$ expressed by Eqs. (74)-(76), 
which are in good agreement with results obtained
in DMF theory shown by solid curves.
If we define the coupling constant $w_m(N)$ 
for which $S_{max}$ is, for example, 0.3 for a given $N$,
we get $w_m(N)$=0.101, 0.147
0.237 and 0.322 for $N$=10, 20, 50 and 100, respectively,
which lead to
%When these values are normalized by $w_m$ for $N=10$,
%they are given by 
$w_m(N)/w_m(10)$= 1.0, 1.46, 2.35 and 3.19, 
%for $N=$ 10, 20, 50 and 100, 
respectively,
when $w_m(N)$ is normalized by $w_m(N=10)$.
This suggests that we may get $w_m(N) \propto \sqrt{N}$.
This arises from the fact that 
the relation: $S_{max} \propto w^2/N$ 
is nearly hold for $S_{max}=0.3$
for which the contribution from the $w^2$ term is
more considerable than that from the $w$ term in Eqs. (74)-(76). 
Of course, it is not the case for much smaller value of
$S_{max}$ for which 
the first term is more dominant than the second term.

Expressions of Eqs. (E1)-(E8) for
$w$- and $N$-dependence of fluctuations and the
synchronization ratio, which are obtained 
based on the results calculated in DMF theory,
are useful in a phenomenological sense.
For example, in the case of $negative$ (inhibitory) 
couplings, Eqs.(E1) and (E8) yield 
an increase in $\delta t_{o\ell}$
and a negative $S$, which are supported
by numerical calculations with DMF theory and 
simulations (not shown).
We have tried to extract coefficients
$a_1$, $a_2$, $b_1$ and $b_2$ in Eqs. (E1)-(E8), 
by expanding Eqs. (20)-(29)
in terms of $w$, but have not succeeded yet. 

\section{Large Cluster consisting of Multiple Sub-clusters}

\subsection{Formulation}

It is possible to extend our DMF theory to
a large FN neuron cluster which is divided 
into multiple $M$ sub-clusters
according to their functions.
Dynamics of a single FN neuron $i$ 
in a given sub-cluster $m$ (=1 to $M$)
which consists of $N_m$ neurons,
is described by the nonlinear DEs given by 
\begin{eqnarray}
\frac{dx_{i}(t)}{dt} &=& F[x_{i}(t)]
- c y_{i}(t) + I_{i}^{(c1)}(t) + I_{i}^{(c2)}(t) 
+I_{m}^{(e)}(t) + \xi_i(t),  \\
\frac{dy_{i}(t)}{dt} &=& b x_{i}(t) - d y_{i}(t)+e,
\;\;\;\mbox{($i=1-N_m$)}
\end{eqnarray}
where 
%$F[x(t)]=k\: x(t)\: [x(t)-a]\: [1-x(t)]$, 
%$k=0.5$, $a=0.1$, $b=0.015$, $d=0.003$ and $e=0$ 
%\cite{Rod96}\cite{Tuckwell98},
$x_i$ and $y_i$ denote the fast (voltage) variable
and slow (recovery) variable, respectively,
$\xi_i(t)$ the Gaussian white noise with $<\xi_i(t)>=0$ and
$<\xi_i(t)\:\xi_j(t')>=\beta_i^2 \; \delta_{ij}\:\delta(t-t')$,
the bracket $< \cdot >$ denoting the average \cite{note2}.
In Eq. (77),
$I_i^{(c1)}(t)$ and $I_i^{(c1)}(t)$ given by
\begin{eqnarray}
I_i^{(c1)}(t)&=&(w_{m m}/N_{m})\;\sum_{j \in m} G(x_{j}(t)), \\
I_i^{(c2)}(t)&=&\sum_{n (\neq m)} (w_{m n}/N_{n})
\;\sum_{k \in n} G(x_{k}(t)),
\end{eqnarray}
express the couplings within the sub-cluster $m$
with the strength $w_{mm}$,
and those between sub-clusters with
the strength $w_{mn}$, respectively,
$N_m$ the number of neurons in the sub-cluster $m$, and
$G(x)$ is the sigmoid function.
$I_m^{(e)}(t)$ stands for an external single spike input 
applied to all neurons in the sub-cluster $m$,
as given by Eq. (4).

As in the Sec. IIA, 
%After Tuckwell and Rodriguez \cite{Tuckwell96}\cite{Tuckwell98},
%we will express these nonlinear differential equations
%by the moments of variables.
we first define the global variables for the sub-cluster $m$ by
\begin{eqnarray}
X^m(t)&=&(1/N_{m})\;\sum_{i \in m} \;x_{i}(t), \\
Y^m(t)&=&(1/N_{m})\;\sum_{i \in m} \;y_{i}(t),
\end{eqnarray}
and their averages by
\begin{eqnarray}
\mu_1^m(t)&=&<X^m(t)>,  \\ 
\mu_2^m(t)&=&<Y^m(t)>,
\end{eqnarray}
%where the bracket $< \cdot>$ denotes the average.
%Next we express the differential equations given by 
%Eqs. (1) and (2) in terms
%of the deviations from their averages defined by
%\begin{eqnarray}
%\delta x_i(t)&=& x_i(t)-\mu_1^{m}(t), \\
%\delta y_i(t)&=& y_i(t)-\mu_2^{m}(t),
%\end{eqnarray}
%to get
%\begin{eqnarray}
%\frac{d x_i}{d t} &=& F(\mu_1^{m}) + F'(\mu_1^{m}) \delta x_i 
%+\frac{1}{2} F^{(2)}(\mu_1^{m}) \delta x_i^2 
%+\frac{1}{6} F^{(3)}(\mu_1^{m}) \delta x_i^3 \nonumber \\
%&& -c \mu_2^m -c \delta y_i + I_i^{(c1)} + I_i^{(c2)} 
%+ I_i^{(e)} + \xi_i, \\
%\frac{d y_i}{d t} &=& b \mu_1^{m} - d \mu_2^{m} 
%+b \delta x_i - d \delta y_i+e,
%\end{eqnarray}
%with
%\begin{eqnarray}
%I_i^{(c1)}&=& w_{m m} \{ (1-\frac{1}{N}) G(\mu_1^{m}) \nonumber \\
%&&+ \frac{1}{N_{m}} \sum_{j (\neq i) \in m} [G'(\mu_1^m) \delta x_j 
%+ \frac{1}{2} G^{(2)}(\mu_1^m) \delta x_j^2
%+ \frac{1}{6} G^{(3)}(\mu_1^m) \delta x_j^3])\\
%I_i^{(c2)} &=&\sum_{n (\neq m)} w_{m n} 
%\{ G(P^{n}) 
%+\frac{1}{N_n} 
%\sum_{k \in n} [G'(P^{n}) \delta x_k 
%+ \frac{1}{2} G^{(2)}(P^{n}) \delta x_k^2
%+ \frac{1}{6} G^{(3)}(P^{n}) \delta x_k^3] \}
%\end{eqnarray}
%where the argument $t$ is hereafter neglected. 
Next we define variances and covariances 
between intra-neuron variables, given by 
\begin{eqnarray}
\gamma^{m}_{1,1}
&=& \frac{1}{N_m}\; \sum_{i \in m} <(\delta x_i^m)^2>, \\
\gamma^{m}_{2,2}
&=& \frac{1}{N_m}\; \sum_{i \in m} <(\delta y_i^m)^2>, \\
\gamma^{m}_{1,2}
&=& \frac{1}{N_m}\; \sum_{i \in m} <(\delta x_i^m \;\delta y_i^m)>, 
\end{eqnarray}
where $\delta x_i^m=x_i(t)-\mu_1^m(t)$
and $\delta y_i^m=y_i(t)-\mu_2^m(t)$,
and those between inter-neuron variables, given by
\begin{eqnarray}
\rho^m_{1,1}&=& <(\delta X^m)^2>, \\
\rho^m_{2,2}&=& <(\delta Y^m)^2>, \\
\rho^m_{1,2}&=& <(\delta X^m \; \delta Y^m)>, 
\end{eqnarray}
where $\delta X^m=X^m(t)-\mu_1^m(t)$
and $\delta Y^m=Y^m(t)-\mu_2^m(t)$.
After some manipulations, 
we get the following differential equations:
\begin{eqnarray}
\frac{d \mu_1^{m}}{d t}&=&f^{m}_0 + f^{m}_2 \gamma^{m}_{1,1} -c \mu_2^m  
+ w_{m m}\;(1-\frac{1}{N_m})\;U_{om}
+ \sum_{n (\neq m)}w_{m n}\;U_{0n}
+I_m^{(e)}(t), \\
\frac{d \mu_2^{m}}{d t}&=& b \mu_1^{m} - d \mu_2^{m} +e,  \\
\frac{d \gamma^m_{1,1}}{d t}&=& 2 [(f^m_1+ 3 f^m_3 \gamma^m_{1,1} ) 
\gamma^m_{1,1}- c \gamma^m_{1,2}] 
+ 2 w_{mm}\;(\rho^m_{1,1}-\frac{\gamma^m_{1,1}}{N_m} )\;U_{1m} \nonumber \\
&& + 2 \sum_{n (\neq m)} w_{mn} \rho^n_{1,1}\;U_{1n}
+\beta_{m}^2, \\
\frac{d \gamma^m_{2,2}}{d t}&=& 2 (b \gamma^m_{1,2}- d \gamma^m_{2,2}),  \\
\frac{d \gamma^m_{1,2}}{d t}&=& b \gamma^m_{1,1}
+ (f^m_1+3 f^m_3 \gamma^m_{1,1}-d) \gamma^m_{1,2} 
- c \gamma^m_{2,2} 
+ w_{mm}(\rho^m_{1,2}-\frac{\gamma^m_{1,2}}{N_m})\;U_{1m} \nonumber \\
&& + \sum_{n (\neq m)} w_{mn} \rho^n_{1,2}\;U_{1n},  \\
\frac{d \rho^m_{1,1}}{d t}&=& 2 [(f^m_1
+ 3 f^m_3 \gamma^m_{1,1} )\rho^m_{1,1} - c \rho^m_{1,2}]
+ 2 w_{mm} \;(1-\frac{1}{N_m})\; \rho^m_{1,1}\;U_{1m}
+ \frac{\beta_m^2}{N_m}, \\
\frac{d \rho^m_{2,2}}{d t}&=& 2 (b \rho^m_{1,2}- d \rho^m_{2,2}),  \\
\frac{d \rho^m_{1,2}}{d t}&=& b \rho^m_{1,1}
+  (f^m_1+3 f^m_3 \gamma^m_{1,1}-d)\rho^m_{1,2} 
- c \rho^m_{2,2}+ w_{mm} \;(1-\frac{1}{N_m})\;\rho^m_{1,2}\;U_{1m},  
\end{eqnarray}
with 
\begin{eqnarray}
U_{0m}&=&g^m_o+g^m_2 \gamma^m_{1,1}, \\
U_{1m}&=&g^m_1+3 g^m_3 \gamma^m_{1,1},
\end{eqnarray}
where
$f^m_{\ell} = (1/\ell !) F^{(\ell)}(\mu_1^m)$,
$g^m_{\ell} = (1/\ell !) G^{(\ell)}(\mu_1^m)$. and
$\beta_m^2 = \frac{1}{N_m}\sum_{i \in m} \; \beta_i^2$.
Now we have to solve $8 M$ dimensional deterministic
differential equations, which is more amenable than
to solve $2 N M$ stochastic DEs.

When a given cluster can be divided into excitatory
and inhibitory sub-groups and when
variance and covariance terms in Eqs. (91)-(98) are
neglected, we get
\begin{eqnarray}
\frac{d \mu_1^{E}}{d t}&=&f^{E}_0 -c \mu_2^E 
+ w_{E E}\;U_{E} + w_{E I}\;U_{I}
+I_{E}^{(e)}(t), \\
\frac{d \mu_2^{E}}{d t}&=& b \mu_1^{E} - d \mu_2^{e} +e, \\
\frac{d \mu_1^{I}}{d t}&=&f^{I}_0 -c \mu_2^I 
+ w_{I I}\;U_{I} + w_{I E}\;U_{E}
+I_{I}^{(e)}(t), \\
\frac{d \mu_2^{I}}{d t}&=& b \mu_1^{I} - d \mu_2^{I} +e, 
\end{eqnarray}
where the supra-script $E$ and $I$ stand for the excitatory and inhibitory
clusters, respectively.
This corresponds to the result of Wilson and Cowan \cite{Wilson73}.
Then our DMF theory given by Eqs. (91)-(98) may be
regarded as a generalized version of the Wilson-Cowan
theory including fluctuations
of local and global variables.

\vspace{0.5cm}
\noindent
{\bf Firing-Time Distribution}

The fraction of firings of neurons
in the sub-cluster $m$ is given by
\begin{equation}
Z_{om}(t)= \phi(\frac{\theta-\mu_1^m}{\sigma_{\ell m}}) \;
\frac{d}{dt}(\frac{\mu_1^m}{\sigma_{\ell m}}) \; \Theta(\dot{\mu_1}^m),
\end{equation} 
with
\begin{equation}
\sigma_{\ell m}=\sqrt{\gamma^m_{1,1}}.
\end{equation}
When we expand $\mu_1^m(t)$ in Eq. (105) around 
$t^{*}_{om}$ where
$\mu_1^m(t^{*}_{om})=\theta$,
it becomes
\begin{equation}
Z_{om}(t) \sim 
\phi(\frac{t-t^{*}_{om}}{\delta t_{om}})\;
\frac{d}{dt}(\frac{\mu_1^m}{\sigma_{\ell m}}) \; \Theta(\dot{\mu_1}^m),
\end{equation} 
with
\begin{equation}
\delta t_{o m}=\frac{\sigma_{\ell m}}{\dot{\mu_1}^m},
\end{equation} 
where $\mu_1^m$, $\dot{\mu_1}^m$ and $\sigma_{\ell m}$ are evaluated 
at $t=t^{*}_{om}$.
This shows that most of firing times of a given sub-cluster $m$ 
locate in the range given
as
\begin{equation}
t_{om} \in [t^{*}_{om}-\delta t_{o m},
\; t^{*}_{om}+\delta t_{o m}].
\end{equation} 

\vspace{0.5cm}
\noindent
{\bf Synchronous Response}

The synchronization ratio of a given sub-cluster $m$ is given by
\begin{equation}
S_m(t)
=\frac{(\rho^m_{1,1}/\gamma^m_{1,1}-1/N_m)}{(1-1/N_m)},
\end{equation}
which is 0 and 1 for completely asynchronous and synchronous states,
respectively.

\subsection{Calculated results}

We have performed model calculations, assuming 
$M\;(=10)$ sub-clusters, each of which consists of
$N_m\;(=N)$ neurons.  They are connected by feed-forward
inter-sub-cluster coupling
given by $w_{mn}=w_2 \;\delta_{n m-1}$,
which is allowed to be different from the intra-sub-cluster
coupling given by $w_{mm}=w_1$ for all $m$.
A single spike input given by Eq. (4) is applied only to the
first sub-cluster ($m=1$), and an output of
a sub-cluster $m$ is  
subsequently forwarded to the next sub-cluster $m+1$.
This is conceptually similar to 
the synfire chain \cite{Abeles93}.
When $w_2$ is too small, signals cannot propagate
through sub-clusters.
The critical value of the inter-sub-cluster coupling $w_{2c}$, below which
a spike cannot propagate through sub-clusters,
is $w_{2c}=$0.064, 0.028 and 0.020 for $w_1$=0.0, 0.1 and 0.2,
respectively, with $\beta=0.0$ and $N=100$.

Figure 8(a) shows the time course of $Z_{om}(t)$ 
calculated in DMF theory 
with $\beta=0.05$, $w_1=w_2=0.1$ and $N=100$.
Signals propagate through sub-clusters with
$\delta t_{o m}\; \simeq\;0.8$ for all $m$.
The result is in good agreement with that
obtained in direct simulations (not shown).
Synchronization ratios $S_m(t)$ shown in Fig. 8(b)
have double peaks [see Figs. 6(a) and 6(b)].
The maximum value of $S_m(t)$, for example, is
0.022 for $m=1$ at $t=122.2$.

In contrast, Fig. 8(c) shows the time course
of $Z_{om}(t)$ for the increased
noise intensity of $\beta=0.23$, which shows 
that signals cannot propagate,
dying out at the sixth sub-cluster.
In this case, the agreement of DMF results
with simulations is not satisfactory. 
Synchronization ratios $S_m(t)$ for $\beta=0.23$ 
shown in Fig. 8(d) have multiple peaks for
$1 \leq m \leq 4$, double peaks for $m=5$, a single peak
for $m=6$, and it disappears for $m > 6$.

Figure 9(a) shows
the $m$-dependence of local fluctuations $\delta t_{o m}$
for various $\beta $ with $w_1=w_2=0.1$ and $N=100$.
We note that 
$\delta t_{o m}$ is almost constant for $\beta$=0.05 and 0.10. 
In the case of $\beta=0.23$, however, $\delta t_{o m}$ 
is divergently increased at $m = 5$.
This behavior is not changed when we adopt
a different set of parameters.
Fig. 9(b) shows a similar plot of $\delta t_{o m}$ 
as a function of $m$ for $w_1=0.0$, $w_2=0.1$ and $N=100$.
Signals propagate with $\delta t_{o m}$= 0.04 and 0.12
for $\beta$=0.01 and 0.05, respectively.
For $\beta=0.09$, however, a spike dies out at $m=8$.

Figure 10 shows the $w_1$ dependence of 
the critical noise strength $\beta_c$ above which
signals cannot propagate. 
We get $\beta_c=0.09$ and 0.23 for $w_1=0.0$ and 0.1, respectively,
for $N=100$ as discussed above.
When $w_1$ is set to be 0.2, $\beta_c$ becomes 0.38 for $N=100$.
We note that $\beta_c$ is almost linearly increased
by increasing $w_1$. 
Figure 10 also shows that 
the critical value of $\beta_c$ becomes larger
as the size of sub-cluster ($N$) is larger.

%\newpage
\section{Conclusion and Discussion}

We have proposed DMF theory for stochastic FN neuron 
ensembles, in which
means, variances and covariances of {\it local}
and {\it global} variables are taken into account.
DMF theory has been shown to be derived in various ways:
series expansions of means, variances and covariances
of local and global variables (Sec. IIA),
re-arrangement of moments in RT's method (Sec. IIB) and 
a single-site approximation to RT's method [Sec. IIC] \cite{note3}.
Our DMF theory, which assumes weak noises 
%(2) weak couplings
and the Gaussian distribution of state variables,
goes beyond the weak coupling because
no constraints are imposed on the coupling strength.
Calculated results based on DMF theory are in fairly good agreement
with those obtained by direct simulations for weak noises.
When the noise intensity becomes stronger,
the state-variable distribution more deviates from the 
Gaussian form (see Fig. 3 of \cite{Tanabe01}), 
and the agreement of results
of DMF theory with those of simulations becomes worse.
%We suppose that the critical noise intensity $\beta_c$
%calculated by DMF theory shown in Fig. 10 may be overestimated.
Nevertheless, our DMF theory is expected to be meaningful
for qualitative or semi-quantitative discussion
on the properties of neuron ensembles or clusters.
It is possible to regard nonlinear differential equations given 
by Eqs. (20)-(27) [or Eqs. (91)-(98)]
as the {\it mean-field FN model} for neuron ensembles or clusters.
We hope that
our DMF theory may play a role of the molecular-field
(Weiss) theory in magnetism \cite{Weiss07}: 
the Weiss theory provides a clear physical picture on various
magnetic properties despite some disadvantages such that
it yields too-high critical (Curie) temperature,
wrong critical indices 
and wrong temperature dependence for 
magnetization at low temperatures.
Our DMF theory may be applied to a general
conductance-based nonlinear systems.
When it is applied to an ensemble
of $N \;$HH neurons,
we get the $24$ deterministic
nonlinear differential equations, which
are more amenable 
than original $4 N$ stochastic equations.
Furthermore our DMF theory
based on means,
variances and covarinces of local and global variables
can be applied to more general stochastic systems
besides neural networks.

In summary,
we have developed a semi-analytical DMF theory for
FN neuron ensembles.
% in which dynamics 
%of $N$ FN neurons is described by
%eight nonlinear differential equations
%for moments of local and global variables.
In order to show the feasibility 
of the DMF theory, we have studied the response of
ensembles of FN neurons to a single spike input.
The result is summarized as follows: 
(i) the spike timing precision
of the global variable is much improved by increasing
the ensemble size, even when there is no couplings
among constituent neurons,
(ii) by increasing the coupling strength, the spike transmission
is enhanced by the synchronous response, and 
(iii) the spike propagation with a fairly precise timing is
possible in large-scale clusters when
the noise strength is moderate.
The origin of the item (i) is the same 
as that yielding the central-limit theorem.
Couplings work to suppress local
fluctuations and to increase the synchronization ratio [Eq. (72)].
Items (i) and (ii) are consistent with the results
reported previously \cite{Pei96a}-\cite{Hasegawa02b}.
The item (iii) agrees with the result of recent
simulations for synfire chains, each layer of which 
consists of 100 IF neurons \cite{Diesmann99}.
Items (i)-(iii) are beneficial to the
population temporal-code hypothesis mentioned
in the introduction.
Although calculations reported in this paper
have been limited to supra-threshold inputs, it is possible to
study the response to sub-threshold inputs with the use of DMF theory.
We may investigate combined effects of
white noises and the heterogeneity in model parameters,
which have been intensively studied in recent years \cite{Hu00}.
Such calculations are in progress and will be
reported in a separate paper.

\section*{Acknowledgements}
The author would like to express his sincere thanks to
Professor Hideo Nitta for critical reading of the manuscript.
This work is partly supported by
a Grant-in-Aid for Scientific Research from the Japanese 
Ministry of Education, Culture, Sports, Science and Technology.

\appendix
\section{Derivation of Eqs. (20)-(27) and Eqs. (32)-(36)}

%\noindent
%{\bf APPENDIX A: Derivation of Eqs. (20)-(27)}

From Eqs. (9)-(12), we get the differential equations for the
deviations of $\delta x_i$ and $\delta y_i$ of the neuron $i$,
given by
\begin{eqnarray}
\frac{d \delta x_i}{d t}&=& f_1 \delta x_i+f_2 (\delta x_i^2-\gamma_{1,1})
+ f_3 \delta x_i^3 - c \delta y_i + \xi_i 
+ \delta I_i^{(c)}, \\
\frac{d \delta y_i}{d t}&=& b \delta x_i - d \delta y_i,
\end{eqnarray}
with
\begin{eqnarray}
\delta I_i^{(c)}
&=&  w \left( \frac{g_1}{N} \sum_{j(\neq i)} \delta x_j
+ g_2 [\frac{1}{N}\sum_{j(\neq i)} \delta x_j^2 
-(1-\frac{1}{N})\gamma_{1,1}]
+  \frac{g_3}{N} \sum_{j(\neq i)} \delta x_j^3 \right), 
\end{eqnarray}
The differential equations for the variances and
covariances are given by
\begin{eqnarray}
\frac{d \gamma_{\kappa,\lambda}}{d t}
&=&\frac{d}{dt} \frac{1}{N} \sum_{i}
< [(\delta x_i^2) \:\delta_{\kappa 1}\delta_{\lambda 1}
+ (\delta x_i \delta y_i)\: \delta_{\kappa 1}\delta_{\lambda 2} 
+ (\delta y_i^2)\: \delta_{\kappa 2}\delta_{\lambda 2}]>,
\nonumber \\
&=& \frac{1}{N} \sum_{i}
< \{ 2 [\delta x_i \:(\frac{d \delta x_i}{d t})] 
\; \delta_{\kappa 1} \delta_{\lambda 1}
+ [\delta y_i \:(\frac{d \delta x_i}{d t})
+\delta x_i \:(\frac{d \delta y_i}{d t})] 
\delta_{\kappa 1} \delta_{\lambda 2} \nonumber \\
&+& 2 [\delta y_i\:(\frac{d \delta y_i}{d t})] 
\; \delta_{\kappa 2} \delta_{\lambda 2} \}>,
\end{eqnarray}

\begin{eqnarray}
\frac{d \rho_{\kappa,\lambda}}{d t}
&=& \frac{d}{dt}
(1/N^2)\;\sum_{i}\;\sum_{j}
<[(\delta x_i \delta x_j) \;\delta_{\kappa 1} \delta_{\lambda 1} 
+ (\delta x_i \delta y_j) \;\delta_{\kappa 1} \delta_{\lambda 2}
+ (\delta y_i \delta y_j) \;\delta_{\kappa 2}\delta_{\lambda 2} ] > \nonumber\\
&=& \frac{1}{N^2} \sum_{i} \sum_{j} 
< \{ 2 [\delta x_i \:(\frac{d \delta x_j}{d t})] 
\; \delta_{\kappa 1} \delta_{\lambda 1} 
+ [\delta y_i \:(\frac{d \delta x_j}{d t})
+\delta x_i \:(\frac{d \delta y_j}{d t})] 
\delta_{\kappa 1} \delta_{\lambda 2} \nonumber \\
&+& 2 [\delta y_i\:(\frac{d \delta y_j}{d t})] 
\; \delta_{\kappa 2} \delta_{\lambda 2} \}>,
\end{eqnarray}
In the process of the calculation using Eqs. (20)-(27),
we have adopted the following approximations:

\noindent
(1) the forth-order variances are assumed to be \cite{Tanabe01}
\begin{eqnarray}
\frac{1}{N} \sum_i <\delta x_4>&=&3\;\gamma_{1,1}\gamma_{1,1}, \\
\frac{1}{N} \sum_i <\delta x_3 \delta y_i>&=&3\;\gamma_{1,1}\gamma_{1,2},
\end{eqnarray}
and
\begin{eqnarray}
\frac{1}{N^2} \sum_{i} \; \sum_{j}
<\delta x_i \delta x_j^3> &=& 3 \gamma_{1,1} \rho_{1,1},  \\
\frac{1}{N^2} \sum_{i} \; \sum_{j}
< \delta y_i \delta x_j^3> &=& 3 \gamma_{1,1} \rho_{1,2},
\end{eqnarray}
other forth terms being set zero.

\noindent
(2) the third-order variances and
terms higher than forth order are neglected,

Calculations of Eqs. (32)-(36) are siimilar to those discussed above
if we read as $m_1^i \rightarrow \mu_1$, 
$m_2^i \rightarrow \mu_2$,   
$\Delta x_i \rightarrow \delta x_i$ and
$\Delta y_i \rightarrow \delta y_i$. 

\section{Some Limiting Cases of Eqs. (20)-(27)}

\noindent
(1) In the limit of a single ($N=1$) neuron, Eqs. (20)-(27) reduce to 
\begin{eqnarray}
\frac{d \mu_1}{d t}&=&f_0 + f_2 \gamma_{1,1} -c \mu_1  + I^{(e)}(t), \\
\frac{d \mu_2}{d t}&=& b \mu_1 - d \mu_2 +e,   \\
\frac{d \gamma_{1,1}}{d t}&=& 2 (f_1 \gamma_{1,1}+3 f_3 \gamma_{1,1}^2 - c \gamma_{1,2}) 
+ \beta^2,  \\
\frac{d \gamma_{2,2}}{d t}&=& 2 (b \gamma_{1,2}- d \gamma_{2,2}),  \\
\frac{d \gamma_{1,2}}{d t}&=& b \gamma_{1,1}
+ (f_1-d) \gamma_{1,2} +3 f_3 \gamma_{1,1}\gamma_{1,2}- c \gamma_{2,2},  \\
\rho_{\kappa,\lambda}&=&\gamma_{\kappa,\lambda}.
\end{eqnarray}
Equations (B1)-(B5) agree with the results of 
Rodriguez and Tuckwell (RT) \cite{Tuckwell98} and 
Tanabe and Pakdaman (TP) \cite{Tanabe01}.
In RT, the forth terms 
which appear in the process of calculating
$d \gamma_{1,1}/d t$ and $d \gamma_{1,2}/d t$ in Eqs. (B3) and (B5),
are assumed to be zero,
whereas in TP, they are assumed to be
as given by Eqs. (A6) and (A7).
\vspace{0.5cm}
\noindent

(2) In the limit of large $N$, where
the exclusion of the self-couplings in Eq. (3) may be
neglected, Eqs. (20)-(27) become

\begin{eqnarray}
\frac{d \mu_1}{d t}&=&f_0 + f_2 \gamma_{1,1} -c \mu_2  
+ w\;U_0+I^{(e)}(t), \\
\frac{d \mu_2}{d t}&=& b \mu_1 - d \mu_2 +e, \\
\frac{d \gamma_{1,1}}{d t}&=& 2 [(f_1+3f_3\gamma_{1,1}) \gamma_{1,1}- c \gamma_{1,2}] 
+ 2 w \rho_{1,1} \:U_1+\beta^2, \\
\frac{d \gamma_{2,2}}{d t}&=& 2 (b \gamma_{1,2}- d \gamma_{2,2}), \\
\frac{d \gamma_{1,2}}{d t}&=& b \gamma_{1,1}+ (f_1+3f_3\gamma_{1,1} -d) \gamma_{1,2} 
- c \gamma_{2,2}
+ w  \rho_{1,2} U_1, \\
\frac{d \rho_{1,1}}{d t}&=& 2 [(f_1+3f_3\gamma_{1,1})  \rho_{1,1} - c \rho_{1,2}]
+ 2 w \rho_{1,1} U_1 + \frac{\beta^2}{N},  \\
\frac{d \rho_{2,2}}{d t}&=& 2 (b \rho_{1,2}- d \rho_{2,2}),  \\
\frac{d \rho_{1,2}}{d t}&=& b \rho_{1,1}+ (f_1+3f_3\gamma_{1,1} -d) \rho_{1,2} 
- c \rho_{2,2}
+ w \rho_{1,2}\;U_{1},  
\end{eqnarray}
where $U_0=g_0+g_2 \gamma_{1,1}$ and $U_1=g_1+3 g_3 \gamma_{1,1}$.

%\vspace{0.5cm}
%\noindent
%(3) In the noise-free case ($\beta=0$), we get
%\begin{eqnarray}
%\frac{d P}{d t}&=&f_0 - c Q  +e
%+ w\;(1-\frac{1}{N})\;U_0 +I^{(e)}(t), \\
%
%\frac{d Q}{d t}&=& b P - d Q +h, 
%\end{eqnarray}
%because $C_{k,\ell} = D_{k,\ell} = 0$. 

%\newpage
\section{Derivation of Eq. (54)}

The distribution $P(x_i)$ in Eq. (54) is formally given by
\begin{equation}
P(x_i) = \int ... \int \; \Pi_{j(\neq i)} \:dx_j \:\Pi_j \:dy_j\;
p(x_1,....,x_N,y_1,....,y_N),
\end{equation}
with the probability distribution function (pdf) of
$p(x_1,....,x_N,y_1,....,y_N)$
for the $2N$-dimensionl vector
${\bf z}=(x_1,....,x_N,y_1,....,y_N)$, given by
\begin{equation}
p(x_1,....,x_N,y_1,....,y_N)
=\frac{1}{(2 \pi)^N \sqrt{\mid {\bf V} \mid}} 
%exp [-\frac{1}{2}Q(x_1,y_1)],
\; exp [-\frac{1}{2} {({\bf z}-\mbox{\boldmath {$\mu$}})}^{t} \;
{\bf V}^{-1} \;{({\bf z}-\mbox{\boldmath {$\mu$}})}],
\end{equation} 
where
{\boldmath $\mu$} and ${\bf V}$ express the mean vector 
and the variance-covariance matrix, respectively,

In the case of a single FN neuron ($N=1$), pdf is given by
\begin{equation}
p(x_1,y_1)=p_1(x_1,y_1)
=\frac{1}{(2 \pi) \sqrt{\mid {\bf V} \mid}} 
%exp [-\frac{1}{2}Q(x_1,y_1)],
\; exp [-\frac{1}{2} {({\bf z}-\mbox{\boldmath {$\mu$}})}^{t} \;
{\bf V}^{-1} \;{({\bf z}-\mbox{\boldmath {$\mu$}})}],
\end{equation} 
with
%\begin{equation}
%Q(x_i,y_i)=a_{11}(x_i-\mu_1)^2
%+2 a_{12}(x_i-\mu_1)(y_i-\mu_2)
%+a_{22}(y_i-\mu_2),
%\end{equation} 
\begin{eqnarray}
{\bf z}&=&(x_1,y_1)^{t}, \\
\mbox{{\boldmath $\mu$}}&=&(\mu_1, \mu_2)^{t},
\end{eqnarray}
%\begin{equation}
%a_{jk}=(V^{-1})_{jk},
%\end{equation} 
\begin{equation}
{\bf V}=
\left(
\begin{array}{cc}
\gamma_{1,1}& \gamma_{1,2} \\
\gamma_{1,2}& \gamma_{2,2} 
\end{array}
\right).
\end{equation} 
Substituting Eqs. (C3)-(C6) to Eq. (C1), we get
\begin{equation}
P(x_1)=\int dy_1 \; p(x_1,y_1) 
= \frac{1}{\sqrt{\gamma_{1,1}}}
\phi \left( \frac{x_1-\mu_1}{\sqrt{\gamma_{1,1}}} \right),
\end{equation} 
where $\phi(x)$ denotes the normal distribution function:
\begin{equation}
\phi(x)=\frac{1}{\sqrt{2 \pi}} 
{\rm exp}(-\frac{x^2}{2}).
\end{equation} 

%In the case of arbitrary $N$, a calculation of $p(x_i)$ 
%given by Eqs. (C1) and (C2) is difficult. 
%For vanishing couplings ($w=0$), pdf is given as the product
%of the two-dimensional pdf for $N=1$, as given by 
%\begin{equation}
%p(x_1,y_1, ...,x_N,y_N)=\Pi_{j=1}^{N} \;p_1(x_j,y_j),
%\end{equation} 
%where $p_1(x_j,y_j)$ is given by Eq. (C3).
%Then we get
%\begin{eqnarray}
%P(x_i) &=& \int... \int \; dy_i \; p_1(x_i,y_i)\;
%\Pi_{j(\neq i)}  \:dx_j\;\Pi_{j(\neq i)} \:dy_j
%\; \Pi_{j(\neq i)}\;p_1(x_j,y_j), \\
%&=& \frac{1}{\sqrt{\gamma_{1,1}}}
%\phi \left( \frac{x_i-\mu_1}{\sqrt{\gamma_{1,1}}} \right),
%\end{eqnarray}
%If couplings are weak,
%Eq. (C10) is expected to be nearly hold. 
%It is noted that effects of couplings on
%$\gamma_{1,1}$ {\it et. al.} come
%through $w$ term in Eqs. (18)-(20).

In the case of arbitrary $N$ under consideration, 
the calculation of $P(x_i)$ may be performed 
within DMF as folllows.
As mentioned in Sec. IIC,
our DMF theory assumes the configuration
in which a {\it single} neuron is embedded
in an effective medium
characterized by $\mu_{\kappa}$, $\gamma_{\kappa,\lambda}$
and $\rho_{\kappa,\lambda}$ [Eqs. (49)-(51)]. 
Thus it is effectively 
the problem of a single neuron in the effective
medium. 
Means ($\mu_{\kappa}$), 
variances ($\gamma_{\kappa,\lambda}$) 
and covariances ($\rho_{\kappa,\lambda}$)
of local variables are determined by Eqs. (20)-(24).
Then the calculation of $P(x_i)$ 
for $N>1$ is the same as that for $N=1$ mentioned above, and
it is given by
\begin{equation}
P(x_i)
=\int dy_1 \; p(x_1,y_1) 
= \frac{1}{\sqrt{\gamma_{1,1}}}
\phi \left( \frac{x_i-\mu_1}{\sqrt{\gamma_{1,1}}} \right).
\end{equation}

\section{Derivation of Eq. (64)}

Equations (20), (21), (25)-(27) form
DEs for means ($\mu_{\kappa}$), 
variances ($\gamma_{\kappa, \lambda}$)
and covariances ($\rho_{\kappa, \lambda}$) for global variables, 
$X$ and $Y$. Then
$P(X)$ in Eq. (64) is given by
\begin{equation}
P(X)= \int dY p(X,Y),
\end{equation} 
with pdf for the two-dimensional vector
${\bf z}=(X,Y)^{t}$ given by
\begin{equation}
p(X,Y)
=\frac{1}{2 \pi \sqrt{\mid {\bf V} \mid}} 
%exp [-\frac{1}{2}Q(X,Y)],
\; exp [-\frac{1}{2} {({\bf z}-\mbox{\boldmath {$\mu$}})}^{t} \;
{\bf {\bf V}}^{-1} \;{({\bf z}-\mbox{\boldmath {$\mu$}})}],
\end{equation} 
with
\begin{eqnarray}
\mbox{{\boldmath $\mu$}}&=&(\mu_1, \mu_2)^{t},
\end{eqnarray}
%\begin{equation}
%Q(X,Y)=a_{11}(X-\mu_1)^2
%+2 a_{12}(X-\mu_1)(Y-\mu_2)
%+a_{22}(Y-\mu_2),
%\end{equation} 
%\begin{equation}
%%a_{jk}=(V^{-1})_{jk},
%\end{equation} 
\begin{equation}
{\bf V}=
\left(
\begin{array}{cc}
\rho_{1,1}& \rho_{1,2} \\
\rho_{1,2}& \rho_{2,2}
\end{array}
\right).
\end{equation} 
Substituting Eqs. (D2)-(D4) to Eq. (D1), we obtain
\begin{equation}
P(X) 
= \frac{1}{\sqrt{\rho_{1,1}}}
\phi \left( \frac{X-\mu_1}{\sqrt{\rho_{1,1}}} \right),
\end{equation} 
where $\phi(x)$ denotes the normal distribution [Eq. (C8)].

Alternatively $P(X)$ is expressed by
\begin{equation}
P(X) = \int ... \int \; \Pi_{i} \:dx_i \:\Pi_i \:dy_i\;
p(x_1,...,x_N, y_1, ...,y_N) \; \delta(X - \frac{1}{N} \sum_i x_i),
\end{equation} 
where $p(x_1, ...,x_N, y_1, ...,y_N)$ 
stands for pdf for $2N$-dimensional vector [Eq. (C2)].
However, a calculation of $P(X)$ based on Eq. (D6) is difficult except
for the no coupling case ($w=0$), for which
pdf is given by
\begin{equation}
p(x_1, ...,x_N, y_1, ...,y_N) = \Pi_j p_1(x_i,y_i),
\end{equation}
$p_1(x_i)$ being pdf for $N=1$ [Eq. (C3)].
Performing integrals with respect to $y_i$ in Eq. (D6)
with Eq. (D7), we get
\begin{equation}
P(X) = \int ... \int \; \Pi_{i} \:dx_i \;
\Pi_i \frac{1}{\sqrt{\gamma_{1,1}}}
\;\phi\left(\frac{x_i-\mu_1}{\sqrt{\gamma_{1,1}}} \right)
\; \delta(X - \frac{1}{N} \sum_i x_i),
\end{equation} 
By using the procedure conventionally used for proofing
the central-limit theorem,
we obtain Eq. (D5) with $\rho_{1,1}=\gamma_{1,1}/\sqrt{N}$
(for $w=0$).
We should note that, a calculation of $P(X)$
based on Eq. (D1) is easier than that based on Eq. (D6)
and that the former is applicable for finite
couplings.

\section{Analysis of noise, coupling and size dependence}

\noindent
(1) {\bf $\delta t_{o\ell}$ and $\delta t_{og}$}

Based on the calculated results of DMF theory, we have
tried to obtain the analytical expression of $\beta$-, $w$-
and $N$-dependence of $\delta t_{o\ell}$ and $\delta t_{og}$.
Figures 3(a) and 3(b) show that 
$\delta t_{o\ell}$ and $\delta t_{og}$ are proportional 
to $\beta$ for weak noises,
for which both $\gamma_{1,1}$ and $\rho_{1,1}$ are proportional to $\beta^2$ 
[see Eqs. (56), (61), (65) and (68)].
From results shown in Figs. 4 and 5, we have obtained
expressions given by
\begin{eqnarray}
\frac{\delta t_{o\ell}(w,N)}{\delta t_{o\ell}(0,1)}
&\sim& 1 -(\frac{1}{2}) (1-\frac{1}{N})^n\;(a_1 w  + a_2 w^2 +..),  \\
\frac{\delta t_{og}(w,N)}{\delta t_{o\ell}(0,1)}
&\sim&  \frac{1}{\sqrt{N}},
\end{eqnarray}
where $n=1$, $\delta t_{o\ell}(0,1)$=2.71, $a_1=7.0$
and $a_2=-11.0$.
The $N$ dependence of $\delta t_{o\ell}$ expressed by
Eq. (E1) with $n=1$ and 2 are shown by thin solid curves 
at the uppermost in Fig. 4(b)
with DMF result (small filled squares):
these results are shifted upward by 0.433 for a clarity of the figure.
The result with $n=1$ is in better agreement with the DMF result
than that with $n=2$.
On the other hand,
bold, dashed curves in Fig. 5(a)
and 5(b) show the $w$ dependence of $\delta t_{o\ell}$ 
for $N=100$ and 10, respectively, expressed by
Eq. (E1) with $n=1$, which is in good agreement with
results of DMF theory shown by filled squares.
This implies from Eqs. (56), (61), (65) and  (68)
that the $w$- and $N$-dependence of
$\gamma_{1,1}$ and $\rho_{1,1}$ evaluated
at $t=t^*_{o}$ where $\mu_1(t^{*}_{o})=\theta$, are given by
\begin{eqnarray}
\frac{\gamma_{1,1}(w,N)}{\gamma_{1,1}(0,1)}
&\sim& 1 -(1-\frac{1}{N})\:(a_1 w  + a_2 w^2 +..),  \\
\frac{\rho_{1,1}(w,N)}{\gamma_{1,1}(0,1)}
&\sim&  \frac{1}{N}.
\end{eqnarray}
Note that $\delta t_{o\ell}(0,1)$ and $\gamma_{1,1}(0,1)$ are
proportional to $\beta$ and $\beta^2$, respectively.

\vspace{0.5cm}
\noindent
(2) {\bf $S_{max}$}

In order to discuss the expression of $\beta$-,
$w$- and $N$-dependent $S_{max}$,
we have analyzed results of $S_{max}$ 
shown in Fig. 7 by
\begin{eqnarray}
S_{max}&=&c_1 w + c_2 w^2 +...,
\end{eqnarray}
to guess how
expansion coefficients of $c_1$ and $c_2$ depend on $N$.
After several tries, we have concluded
that the $w$- and $N$-dependence of
$\gamma_{1,1}$ and $\rho_{1,1}$ evaluated
at $t=t^{(m)}_{o}$ where $\rho_{1,1}(t)$ has the
maximum value, may be given by
\begin{eqnarray}
\frac{\gamma_{1,1}(w,N)}{\gamma_{1,1}(0,1)}
&\sim& 1 -(1-\frac{1}{N})^{m} \;(b_1 w  + b_2 w^2 +..), \\
\frac{\rho_{1,1}(w,N)}{\gamma_{1,1}(0,1)}
&\sim&  \frac{1}{N},
\end{eqnarray}
yielding $S_{max}$ given by [see Eq. (72)]
\begin{eqnarray}
S_{max}(w,N)=(\frac{1}{N})(1-\frac{1}{N})^{m-1}
\;\{ b_1 w + [b_2 +  (1-\frac{1}{N})^2 b_1^2]\:w^2 \},
\end{eqnarray}
where $m=2$, $b_1=22$ and $b_2=-290$.
Bold, dashed curves in Fig. 7 show
the $w$ dependence of $S_{max}$ expressed by Eq. (E8)
for various $N$ values, which are 
in fairly good agreement with results of DMF theory
shown by solid curves.
We should point out that 
a factor of $(1-1/N)$ in Eqs. (E1), (E3),
(E6) and (E8) appears because
the coupling $w$ does not work 
in a single-neuron case ($N=1$)
and that at least the second power ($m=2$) is necessary in 
Eq. (E6) for $S_{max}$ to vanish in the $N=1$ limit.
A functional form of Eq. (E3) 
may be different from that of Eq. (E6)
because the former is evaluated 
at $t_o^{*}$ while the latter at $t_o^{(m)}$.
%Although we have tried to get analytical expressions as given by
%Eqs. (C3), (C4), (C6) and (C7) 
%by expanding Eqs. (20)-(29) in terms of $w$ 
%and $N$, we have not succeeded yet.
Our DMF calculation shows that when $\beta$ is increased
for a fixed (finite) $w$ value, $S_{max}$ is gradually decreased,
although Eq. (E8) has no $\beta$ dependence.
This is due to contributions of $O(\beta^4)$ to
$\gamma_{1,1}$ and $\rho_{1,1}$, which have been not
included in the above discussion.
%Although we have tried to get analytical expressions as given by
%Eqs. (C3), (C4), (C6) and (C7) 
%by expanding Eqs. (16)-(25) in terms of $w$ 
%and $N$, we have not succeeded yet.

%\begin{references}

%\end{references}

\begin{figure}
\caption{
%Fig B. 
Time courses of means, variances and covariances
calculated by DMF theory (solid curves) and simulations (dashed curves): 
(a) $\mu_1$, (b) $\mu_2$, (c) $\gamma_{1,1}$, (d) $\rho_{1,1}$, 
(e) $\gamma_{2,2}$, (f) $\rho_{2,2}$, (g) $\gamma_{1,2}$ and 
(h) $\rho_{1,2}$,
for $A=0.10$, $\beta=0.01$, $w=0.0$ and $N=100$. 
Results of (d), (f) and (h) are multiplied 
by a factor of hundred. 
The chain curve at the bottom of (a) expresses
a single input spike, $I^{(e)}$ in Eq. (4) [see also Fig. 2(a)].
}
\label{fig1}
\end{figure}

\begin{figure}
\caption{
%Fig. J. 
Time courses of (a) $I^{(e)}$, 
(b) $W_{\ell}$ (the dashed curve) 
and $Z_{\ell}$ (the solid curve) in DMF theory,
(c) $Z_{\ell}$ in simulations, 
(d) $W_{g}$ (the dashed curve) 
and $Z_{g}$ (the solid curve) in DMF theory, and
(e) $Z_{g}$ in simulations, 
for $A=0.10$, $\beta=0.01$,
$w=0.0$ and $N=100$.
}
\label{fig2}
\end{figure}

\begin{figure}
\caption{
%Fig. D. 
(a) The $\beta$ dependence of $\delta t_{o\ell}$ (squares),
and $\delta t_{og}$ (circles) for $w=0.0$ and
(b) that for $w=0.2$ with $N=100$,
filled symbols denoting results in DMF theory
and open symbols those in simulations.
}
\label{fig3}
\end{figure}

\begin{figure}
\caption{
%Fig. E.
Log-log plots of $\delta t_{o\ell}$ (squares)
and $\delta t_{og}$ (circles) against $N$ for (a) $w=0.0$
and (b) $w=0.2$,
filled symbols denoting results in DMF theory
and open symbols those in simulations.
Shown at the uppermost in (b) are the DMF result (small, filled squares) 
and results with Eq. (E1) 
with $n=1$ and 2 
(thin solid curves): they are shifted upward 
by 0.433 for a clarity of the figure (see text).
}
\label{fig4}
\end{figure}

\begin{figure}
\caption{
%Fig. F
The $w$ dependence of $\delta t_{o\ell}$ (squares)
and $\delta t_{og}$ (circles) for (a) $N=100$
and (b) $N=10$ with $\beta=0.01$,
filled symbols denoting results in DMF theory
and open symbols those in simulations.
Bold, dashed curves for $w \leq 0.2$ express Eq. (73) (see text).
}
\label{fig5}
\end{figure}

\begin{figure}
\caption{
%Fig. G. 
The time course of synchronization ratio $S$
for (a) $w=0.1$ and (b) $w=0.2$
with $\beta=0.01$ and $N=100$,
solid curve denoting results of DMF theory and 
dashed curve those of simulations.
}
\label{fig6}
\end{figure}

\begin{figure}
\caption{
%Fig. H.
The $w$ dependence of the maximum of $S$, $S_{max}$,
for $N=10$ (squares), $N=20$ (triangles),
$N=50$ (inverted triangles) and $N=100$ (circles) 
with $\beta=0.01$, filled symbols denoting 
results of DMF theory and open symbols those of simulation.
Bold, dashed curves for $w \leq 0.2$ express Eq. (74) (see text).
}
\label{fig7}
\end{figure}

\begin{figure}
\caption{
%Fig. W.
Time courses of $Z_{om}$ for (a) $\beta=0.05$
and (b)$\beta=0.23$ with $w_1=w_2=0.1$, and
time courses of $S_{m}$ for (c) $\beta=0.05$
and (d)$\beta=0.23$ with $w_1=w_2=0.1$,
calculated for
$N=100$ and $M=10$ by DMF theory.
}
\label{fig8}
\end{figure}

\begin{figure}
\caption{
%Fig. X. 
$\delta t_{o m}$ as a function of $m$
for (a)  $w_1=w_2=0.1$ and 
(b) $w_1=0.1$ and $w_2=0.0$
with $N=100$ and $M=10$.
}
\label{fig9}
\end{figure}

\begin{figure}
\caption{
%Fig. Z. 
The $w_1$ dependence of $\beta_c$
for various $N$ values with $w_2=0.1$.
}
\label{fig10}
\end{figure}


\begin{thebibliography}{99}
\bibitem{note3}
DMF theory may be alternatively derived from RT's moment
method \cite{Rod96} if we adopt the approximation given by
$m_{\kappa}^i = m_{\kappa}$ and
$C_{\kappa,\lambda}^{i,j} = \delta_{i j} \; C_{\kappa,\lambda}
+ (1 - \delta_{i j}) \; D_{\kappa,\lambda}$,
%($\kappa,\lambda=1,2; i,j=1 to N$),
with which Eqs. (46)-(48) lead to
$\mu_{\kappa} = m_{\kappa}$, 
$\gamma_{\kappa,\lambda} =  C_{\kappa,\lambda}$ and
$\rho_{\kappa,\lambda} = \frac{1}{N} \; C_{\kappa,\lambda}
+ (1 - \frac{1}{N}) \; D_{\kappa,\lambda}$.
Then Eqs. (37)-(41) yield DEs for $\mu_{\kappa}$, 
$\gamma_{\kappa,\lambda}$ and 
$\rho_{\kappa,\lambda}$ which are given by Eqs. (20)-(27)
in DMF theory.
\end{thebibliography}
\end{document}